 \documentclass[pra,aps,preprint,showpacs,articles]{revtex4}
 \usepackage{epsf,epsfig}
 \usepackage[psamsfonts]{amssymb}
 \usepackage{amsmath}

 \newcommand{\ket}[1]{|#1\rangle}
 \newcommand{\bra}[1]{\langle #1|}

 \newcommand{\expect}[3]{\langle#1|#2|#3\rangle}
 
 \newcommand{\half}{\mbox{$\textstyle \frac{1}{2}$}}

\begin{document}
\title{ Entanglement Transfer via XXZ Heisenberg chain with DM Interaction}
\author{ Morteza Rafiee$^1$ \footnote{m.rafiee178@gmail.com}, Morteza Soltani$^2$\footnote{msoltani@Phys.ui.ac.ir}, Hamidreza
Mohammadi$^3$ \footnote{h.mohammadi@phys.ui.ac.ir} and Hossein
Mokhtari$^4$ \footnote{hmaftab@yahoo.com} }
\affiliation{$^{1,4}$Department of Physics, University of Yazd,
Yazd, Iran} \affiliation{$^{2,3}$Department of Physics, University
of Isfahan, Hezar Jarib Ave., Isfahan, Iran}
\begin{abstract}
The role of spin-orbit interaction, arises from the
Dzyaloshinski-Moriya anisotropic antisymmetric interaction, on the
entanglement transfer via an antiferromagnetic XXZ Heisenberg
chain is investigated. From symmetrical point of view, the XXZ
Hamiltonian with Dzyaloshinski-Moriya interaction can be replaced
by a modified XXZ Hamiltonian which is defined by a new exchange
coupling constant and rotated Pauli operators. The modified
coupling constant and the angle of rotations are depend on the
strength of Dzyaloshinski-Moriya interaction. In this paper we
study the dynamical behavior of the entanglement propagation
through a system which is consist of a pair of maximally entangled
spins coupled to one end of the chain. The calculations are
performed for the ground state and the thermal state of the chain,
separately. In both cases the presence of this anisotropic
interaction make our channel more efficient, such that the speed
of transmission and the amount of the entanglement are improved as
this interaction is switched on. We show that for large values of
the strength of this interaction a large family of XXZ chains
becomes efficient quantum channels, for whole values of an
isotropy parameter in the region $-2\leq\Delta\leq2$.
\end{abstract}
\maketitle
\section{Introduction}
Recently transmitting a quantum state is a most important task in
quantum information and computation processing \cite{bennet}.
Today this purpose could be achieved in two manner: i) using
standard teleportation protocols and, ii) information transfer via
quantum networks. Bayat et.al \cite{bayat} have shown that the
fidelity of transmission is the same for both cases. Additionally,
the former case employs the flying qbits as quantum channel and
hence the fidelity of transmission reduces due to incompleteness
of stationary-to-flying qbits conversion process \cite{vincenzo}.
Thus the later case is superior to the former, particularly in
solid state devices. Among the numerous quantum systems suitable
for quantum networks implementation, the spin chains offer a great
advantages. One of the most interesting art of the spin chains is
their ability to use as quantum wires in the information transfer
protocols over the short distances
\cite{bose,eisert,giovannetti,osborne,bayat3,Burgarth,Burgarth2,bayat2}.
Tunable spin interaction in these systems plays the key role to
motivate one for using this permanently potential in the quantum
information transfer processing. The major works on spin chain are
in the Ferromagnetic(FM) phase \cite{osborne} and the effects of
temperature \cite{bayat3} and decoherence \cite{Burgarth} have
been investigated for FM channels. There are lesser works on
Antiferromagnetic(AFM) phase \cite{asoudeh,eckert}, but as
mentioned in Refs. \cite{bayat2,bayat}, AFM spin chain have higher
ability in transfer of information and thus is a better and faster
alternative. So we prefer to study AFM spin chain. Fortunately,
Antiferromagnetic spin chains with short length(up to 10 spins)
have been built experimentally \cite{Hirjibehedin}. However, much
attention has been paid to the entanglement in spin systems with
only spin-spin interaction and spin-orbit interaction has been
leaved. The spin-orbit interaction produces and anisotropic part
of exchange interaction between localized conduction band
electrons in crystals with lack of inversion symmetry, including
all low dimensional structures and also bulk semiconductors with
zinc-blende and wurtzite type of crystal lattice \cite{kavokin}.
The main part of the anisotropic interaction has the form of
Dzyaloshinski-Moriya interaction (DMI)
\cite{Dzyaloshinski,Moriya1,Moriya2} which have explained the weak
ferromagnetism of antiferromagnetic crystals
($\alpha-Fe_2O_3,MnCO_3$ and $CrF_3$). As was shown
\cite{hamid,hamid1,wang} in the two-qubit Heisenberg systems the
(DMI) plays an important role. The (DMI) express by
\begin{equation}
\vec{D}.[\vec{S}_1\otimes\vec{S}_2].\nonumber
\end{equation}
This interaction arises from extending the Anderson's theory of
superexchange interaction by including the spin-orbit coupling
effects\cite{Moriya1}. In this paper we interested to consider the
model with DMI and also investigated the effects of this
interaction on the information transfer processing. \\Following
the approach proposed by S. Bose\cite{bose}, we place a spin
encoding the state at one end of the chain (which is now equipped
with DMI) and wait for specific amount of time to let this state
propagate to the other end. Therefore, entanglement could be
transferred from one end of the chain to the other. We will show
that the XXZ chain with DMI can be reduced to the modified XXZ
chain with new coupling constant and rotated Pauli operators. The
modified coupling constant and the angle of rotations are depend
on the strength of DMI. Consequently, the entanglement transfer
protocol through the XXZ+DMI chains becomes the same as a protocol
via a general form of XXZ chain with new definition of coupling
constant and Pauli matrices, but it should be noted that at this
case the initial state may be changed. Indeed, both modified
coupling constant and the angle of rotations, which is referred by
"phase factor" through the text, are the efficient parameters for
entanglement transfer. For the sake of clarity, the effects of
phase factor and coupling constant have been separated. Our
calculation shows that this phase factor does not affect on the
entanglement transfer when $\Delta=0$ for the XXZ chain. Whereas,
this phase factor has desirable effects on the parameters of
entanglement transfer for the case of $\Delta<0$ and undesirable
effects when $\Delta>0$. However, there is a contest between the
phase factor and coupling constant. For large amounts of the
strength of DMI($D$), the effects of coupling constant are more
dominant than the effects of phase factor and has desirable
effects. Also, in the whole range of $\Delta$, the effects of the
modified coupling constants and the phase factor simultaneously
investigated on our main goal, i.e, information or entanglement
transmission. Furthermore, the other advantage of DMI is that the
speed of information transmission increases as increasing the
strength of this interaction. Also, at nonzero temperature this
interaction (DMI) improves both $E_{max}$ and $t_{opt}$ as
increasing the strength of DMI ($D$) for both positive and
negative amounts of $\Delta$.\\
The paper is organized as follows. In Sec. II we introduce the XXZ
Hamiltonian with DMI and discuss the way to change the form of
Hamiltonian to the general XXZ Hamiltonian. The new set of
eigenstates have been introduced in this Sec. In Sec. III we
introduce our state transmission. In Sec.
 IV we show analytically the dependence of concurrence on $D$ and
pase factor. In Sec. V we show our numerical calculation of
entanglement transfer and corresponding times in both zero and
nonzero temperature and calculated the speed of information
transfer while, we summarize our results in Sec. VI.
\section{HAMILTONIAN AND MODEL}
The Hamiltonian of the open XXZ chain in presence of spin-orbit
interaction is defined as
\begin{equation}\label{1}
H_{ch}=\sum\limits_{i=1}^{N_{ch}-1}\{J[\sigma_i^x\sigma_{i+1}^x+\sigma_i^y\sigma_{i+1}^y]+\Delta\sigma_i^z\sigma_{i+1}^z\}
+\sum\limits_{i=1}^{N_{ch}-1}\{\vec{D}.(\vec\sigma_i\times\vec\sigma_{i+1})\},
\end{equation}
Where $N_{ch}$ is the number of spin in $1D$ chain,
$\vec\sigma_i$=($\sigma_i^x$,$\sigma_i^y$,$\sigma_i^z$) is the
vector of pauli matrices, J is the exchange coupling, the
parameter $\Delta$ is the anisotropy exchange coupling in the z
direction and D is the Dzyaloshinski-Moriya vector. Different
phases of chain is depending on different range of J and $\Delta$.
For the case of $J<0$ the chain is called Ferromagnetic(FM)
Heisenberg chain and the case $J>0$ is Antiferromagnetic(AFM)
Heisenberg chain. The AFM Heisenberg chain includes FM
phase($\frac{\Delta}{J}<0$), AFM phase($0<\frac{\Delta}{J}<1$) and
N\'{e}el phase($\frac{\Delta}{J}>1$)\cite{14}. If we take
$\vec{D}=D\hat{z}$, then the above Hamiltonian can be written as
\begin{equation}\label{2}
H_{ch}=\sum\limits_{i=1}^{N_{ch}-1}\{J[\sigma_i^x\sigma_{i+1}^x+\sigma_i^y\sigma_{i+1}^y]+\Delta\sigma_i^z\sigma_{i+1}^z\}
+D\sum\limits_{i=1}^{N_{ch}-1}\{\sigma_i^x\sigma_{i+1}^y-\sigma_i^y\sigma_{i+1}^x\}.
\end{equation}
This Hamiltonian is invariant under z-axis rotation, i.e,
$[H,S_z]=0$, where
$S_z=\half{\sum\limits_{i=1}^{N_{ch}}\sigma_i^z}$. By this
property the Hamiltonian can be express in the form of usual XXZ
model without explicit DMI. For this purpose, the pauli operators
in x-y directions are manipulated by the unitary transformation
which is depend on the spin sites\cite{15}
\begin{eqnarray}\label{3}
U_{N_{ch}}&=&\exp^{-i\sum\limits_{m=2}^{N_{ch}}(m-1)\phi\sigma_m^z},
\end{eqnarray}
where $\phi=tang^{-1}(D{/}J)$. Hence these spin coordinate
transformations reads
\begin{eqnarray}\label{4}
\tilde{\sigma}_i^x&=&\sigma_{i}^x\cos{\phi_i}+\sigma_i^y\sin{\phi_i},\nonumber\\
\tilde{\sigma}_{i}^y&=&-\sigma_{i}^x\sin{\phi_i}+\sigma_i^y\cos{\phi_i},\nonumber\\
\tilde{\sigma}_i^z&=&\sigma_{i}^z,
\end{eqnarray}
where $\phi_i=(i-1)\phi$. So, the modified Hamiltonian is express
as
\begin{eqnarray}\label{5}
\tilde{H}_{ch}&=&U_{N_{ch}}H_{ch}U^\dag_{N_{ch}}=\tilde{J}\sum\limits_{i=1}^{N_{ch}-1}\{\tilde{\sigma}_i^x\tilde{\sigma}_{i+1}^x+
\tilde{\sigma}_i^y\tilde{\sigma}_{i+1}^y\}
+\Delta\sum\limits_{i=1}^{N_{ch}-1}\sigma_i^z\sigma_{i+1}^z,\\
\tilde{J}&=&sgn(J)\sqrt{J^2+D^2}.\nonumber
\end{eqnarray}
The eigenstates of the new Hamiltonian ($\ket{\tilde{\psi_n}}$)
are related to the earlier one ($\ket{\psi_n}$) via
$\ket{\tilde{\psi_n}}=U_{N{ch}}\ket{\psi_n}$. As the model of the
system is specified, we can investigate the information
transmission processing in this system.
\section{Entanglement Transmission}
The quantum information transmission of one part of a two-spin
maximally entangled state ($0^\prime0$) via XXZ+DMI spin chain is
investigated while the spin chain is in its ground state
($\ket{\psi_{gs}}_{ch}$). At t=0 we interact the spin $0$ with the
first spin of the chain. We suppose that the chain is prepare in a
unique grand state, initially. The preparation could be performed
by applying a small magnetic field, if it is required.
Furthermore, the interaction between spin $0$ and first spin of
the chain has the same form of the rest of interaction,
\begin{equation}\label{6}
H_{I}=J(\sigma_0^x\sigma_{1}^x+\sigma_0^y\sigma_{1}^y+\Delta\sigma_0^z\sigma_{1}^z)
+D(\sigma_0^x\sigma_{1}^y-\sigma_0^y\sigma_{1}^x).
\end{equation}
Indeed, the system is consist of $0^\prime0$ and $N_{ch}$ spins
and hence the total length of the system is $N=N_{ch}+2$. The
initial state of the system is
\begin{equation}\label{7}
\ket{\psi(0)}=\ket{\psi^-}_{0^\prime0}\otimes\ket{\psi_{gs}}_{ch},
\end{equation}
where
\begin{equation}\label{8}
\ket{\psi^-}_{0^\prime0}=\frac{\ket{01}-\ket{10}}{\sqrt{2}}.
\end{equation}
This $\ket{\psi(0)}$ is used as a channel which transfer the
entanglement. Therefore, the total Hamiltonian being
\begin{equation}\label{9}
H=I_{0^\prime}\otimes(H_{ch}+H_I),
\end{equation}
By this Hamiltonian, the initial state evolves to the state
$\ket{\psi(t)}=e^{-iHt}\ket{\psi(0)}$ and the two sites reduced
density matrix of can be computed as
$\rho_{mn}(t)=tr_{\widehat{mn}}\{\ket{\psi(t)}\bra{\psi(t)}\}$, where
$tr_{\widehat{mn}}$ is the partial trace over the system except sites m and
n. The two sites reduced density matrix in computational
basis ($\ket{00},\ket{01},\ket{10},\ket{11}$) has the general form
as\cite{hamid}
\begin{equation}\label{10}
\rho_{mn}(t) =\left(\begin{array}{*{20}c}
   {{a(t)}} & {{0}} & {{0}} & {{0}}  \\
   {{0}} & {{x(t)}} & {{z(t)}} & {{0}}  \\
      {{0}} & {{z^*(t)}} & {{y(t)}} & {{0}}  \\
   {{0}} & {{0}} & {{0}} & {{b(t)}}  \\
\end{array}\right).
\end{equation}
Although, the density matrix has been written in the Schr$\ddot{\text o}$dinger
picture, but $\rho_{mn}(t)$ in terms of spin-spin correlation
function could be expressed in the Heisenberg picture as
follows\cite{16}
\begin{eqnarray}\label{11}
a(t)&=&1+\half \langle\sigma_m^z(t)+\sigma_n^z(t)\rangle+\langle\sigma_m^z(t)\sigma_n^z(t)\rangle,\nonumber\\
x(t)&=&1+\half\langle\sigma_m^z(t)-\sigma_n^z(t)\rangle-\langle\sigma_m^z(t)\sigma_n^z(t)\rangle,\nonumber\\
y(t)&=&1-\half\langle\sigma_m^z(t)-\sigma_n^z(t)\rangle-\langle\sigma_m^z(t)\sigma_n^z(t)\rangle,\\
b(t)&=&1-\half\langle\sigma_m^z(t)+\sigma_n^z(t)\rangle+\langle\sigma_m^z(t)\sigma_n^z(t)\rangle,\nonumber\\
z(t)&=&\langle\sigma_m^x(t)\sigma_n^x(t)\rangle+\langle\sigma_m^y(t)\sigma_n^y(t)\rangle+
i(\langle\sigma_m^x(t)\sigma_n^y(t)\rangle-\langle\sigma_m^y(t)\sigma_n^x(t)\rangle),\nonumber
\end{eqnarray}
these correlations are computed in terms of the initial
state (Eq.\ref{7}) and $\sigma_m^{\alpha}(t)=e^{-i H
t}\sigma_m^{\alpha}e^{-i H t}$ where $\alpha=\{x,y,z\}$. Since the
concurrence is directly defined in terms of the density matrix and
so any minimization procedure is not necessary, the concurrence is
used as a measure of entanglement for arbitrary mixed state of two
qubits\cite{17},
\begin{eqnarray}\label{12}
C&=&max\{0,2\lambda_{max}-tr\sqrt{R}\},\\
R&=&\rho\sigma^y\otimes\sigma^y\rho^*\sigma^y\otimes\sigma^y,
\end{eqnarray}
where $\lambda_{max}$ is the largest eigenvalues of the matrix
$\sqrt{R}$. For our density matrix the concurrence results to be
\begin{equation}\label{13}
C=2max\{0,C^{(1)},C^{(2)}\},
\end{equation}
where $C^{(1)}=-\sqrt{xy}$ and $C^{(2)}=|z|-\sqrt{ab}$. Because,
$C^{(1)}$ is always negative here the concurrence is\cite{14}
\begin{equation}\label{14}
C=2max\{0,|z|-\sqrt{ab}\}.
\end{equation}
For our main goal the first site refers to $0^\prime$ and other
site refers to the spin located at the end of the chain, say (j).
So in terms of density matrix the subscript m is change to
$0^\prime$ and n is change to j. The singlet fraction of the
state, $\rho_{0'j}$, as an indicator of the average fidelity of
state transferring could be obtain as
\begin{eqnarray}\label{frac}
F=\expect{\psi^-}{\rho_{0'j}}{\psi^-}=\half(x+y-2z).
\end{eqnarray}
With the aid of $\tilde{H}_{ch}=U_{N_{ch}}H_{ch}U^\dag_{N_{ch}}$
we have
\begin{eqnarray}\label{20}
\langle\sigma_m^\alpha(t)\rangle&=&\langle e^{-i H
t}\sigma_m^\alpha e^{i H t}\rangle\nonumber\\
&=&\bra{0^\prime0}\bra{\psi_{gs}}_{ch}U_{N_{ch}+1}e^{-i\tilde{H}t}\tilde{\sigma}_m^\alpha
e^{i\tilde{H}t}U_{N_{ch}+1}^\dag\ket{\psi_{gs}}_{ch}\ket{0^\prime0},
\end{eqnarray}
with use of the $\ket{\tilde{\psi_n}}=U_{N{ch}}\ket{\psi_n}$, the
expectation value can be express as
\begin{eqnarray}\label{22}
\langle\sigma_m^\alpha(t)\rangle&=&\bra{0^\prime0}\bra{\tilde{\psi}_{gs}}_{ch}U_{N_{ch}}U_{N_{ch}+1}\tilde{\sigma}_m^\alpha(t)
U_{N_{ch}+1}^\dag
U_{N_{ch}}^\dag\ket{\tilde{\psi}_{gs}}_{ch}\ket{0^\prime0},
\end{eqnarray}
where
$U_{N_{ch}}U_{N_{ch}+1}=\exp^{-i\sum\limits_{m=1}^{N_{ch}}(2m-1)\phi\sigma_m^z}$
is phase factor which modify the states on the right hand of above
equation. So, we can conclude that this model is similar to usual
XXZ model with the new strength coupling in XY direction ($\tilde
J$) and the new set of states which are multiplied by the phase
factor. In the following, the effects of these parameters (phase
factor and $\tilde J$) will investigated on entanglement transfer
processing.
\section{Analytical Calculation}
To more clarifying, the concurrence between the $0^\prime$ site and
the end of the chain with the length of($N_{ch}=2$) have been
calculated analytically in the appendix. In these calculations the
explicit form of
$\rho_{mn}(t)=tr_{mn}\{\ket{\psi(t)}\bra{\psi(t)}\}$ was used and
these results are quality compatible with the numerical results
for higher N. From the relation(\ref{23}) the concurrence for the
case of ($\Delta=0$) is
\begin{eqnarray}\label{24}
C(t)&=&\frac{1}{8}\big{(}4\big{|}\cos{(\xi t)}-1\big{|}-\big{|}e^{2i\phi}(1+\cos{(\xi t)})\nonumber\\
&+&i\sqrt{2}\sin{(\xi t)}\big{|}\times\big{|}1+\cos{(\xi
t)}+i\sqrt{2}e^{2i\phi}\sin{(\xi t)}\big{|}\big{)},
\end{eqnarray}
where $\xi=2\sqrt{2}\tilde{J}$ and we have the maximum
entanglement(C=1) for $\xi t=\pi$. As we can see, amounts of
entanglement at the first peak ($E_{max}$) and corresponding
time ($t=t_{opt}$) in this case are independent on the $\phi$.

Furthermore, for the case of $\Delta\neq0$ the concurrence is
obtained in Eq.(\ref{25}). In this case $E_{max}$ and $t_{opt}$
depend on $\phi$ as indicated in Fig. 1. In these figure, for
clarifying the role of $\phi$, individually, we fixed the size of
$\tilde J$. The results show that, in the case of $\Delta<0$,
increasing $\phi$ improves both $E_{max}$ and $t_{opt}$ and for
the case $\Delta>0$, the increase of $\phi$ has undesirable
effects on both $E_{max}$ and $t_{opt}$. The calculation becomes
more involved when $N$ exceeds 4, this prevents one from writing
an analytical expression for the concurrence. So, we solve this
problem numerically.
\section{Numerical Calculation}
\subsection{Entanglement at T=0}
Whereas, in the XXZ spin chain the AFM Heisenberg chains ($J>0$)
is a better candidate for information transfer than FM chains
($J<0$) \cite{bayat2}, we confined our calculations to the case
$J>0$ and take $J=1$ to simplify the calculation. The entanglement
is calculated between $0^\prime$ and spin located at the end of
the chain which has the length of N=8. In Fig. 2(a), $E_{max}$, as
measured by concurrence, have been plotted in the domain of
($-2\leq\Delta\leq2$) for different values of $D$ and Fig. 2(b)
illustrates the behavior of $E_{max}$ as a function of $\Delta$
for different values of $\phi$, where $\tilde J$ is fixed. For the
special case D=0 (i.e, $\tilde J=J$ and $\phi=0$) the results are
the same as in Ref. \cite{bayat}, qualitatively. In this case
$E_{max}$ vanishes at the quantum phase transition (QPT) point
($\frac{\Delta}{J}=-1$) \cite{14}. In the presence of $D$, $\tilde
J\neq J$ and hence the QPT point shifts to the
$\frac{\Delta}{\tilde J}=-1$, e.g, for the case of $D=1$ the QPT
occurs at $\Delta=-\sqrt 2$ \footnote{It is important to note
that, in the presence of DM interaction the phase diagram of the
chain may be changed \cite{langari} and hence we must be care on
employing the words such that phase transition and so on. Whereas,
in this paper, we modeled the XXZ chain with DM interaction by
modified with XXZ chain and hence we can still borrow this words
from the phase transition terminology.}. Despite to the results of
\cite{bayat}, at this modified transition point $E_{max}$ does not
vanish, this is due to the presence of the phase factor $\phi$ as
indicated in Fig. 2(b). Furthermore, the amount of $\tilde J$ is
so large for large values of $D$ and so the QPT point disappear in
the frame of $-2\leq\Delta\leq2$. In this range of $\Delta$ and
$D$, the amount of $E_{max}$ is unsensible to the values of $\phi$
and hence $\tilde J$ plays the main role to quantifying $E_{max}$.
Since, $\frac{\Delta}{\tilde J}$ approaches zero as $D$ becomes
large, the system treat as the case $\Delta=0$, i.e,
$E_{max}\longrightarrow E_{max}(\Delta=0)$.
 Fig. 3(a) reveals that the behavior of $t_{opt}$ is compatible
with the result of Fig. 2(a). This figure shows that $t_{opt}$
decreases as increasing $D$, hence the presence of DMI enhances
the speed of information transmission. Also, Figs. 2(b) and 3(b)
show that the effect of $\phi$ on $E_{max}$ and $t_{opt}$ is
undesirable for the case of $\Delta>0$. In contrast, for the case
$\Delta<0$, the maximum entanglement at the first peak and speed
of transmission enhances with $\phi$. In summary, all of the
chains with $-2\leq\Delta\leq2$ can be used as protocol for
information transfer processing with the same cost.

Fig. 4 depicts the singlet fraction in terms of $\Delta$ for
different values of $D$. The result of this picture confirms the
above mentioned effects of DMI on entanglement transfer
properties.

In order to better illustrate the effects of $D$ on the
entanglement transfer properties, $E_{max}$ is plotted in terms of
$D$ and $\phi$ in Figs. 5(a)-5(d) for different values of $\Delta$
respectively. The dips appearing in Figs. 5(a) and 5(c) are due to
the effects of the phase factor ($\phi$) which is obvious from
Figs. 5(b) and 5(d). The effects of DMI on speed of state
transmission are plotted in Figs. 6(a) and 6(b), more obviously.
\subsection{Thermal Entanglement}
Since, preparing the system at $T=0$ is far from access, the
presence of thermal excitations is unavoidable. So, we consider
the state of channel as a thermal state instead of ground state.
The thermal state of channel at temperature $T$ is given by
density matrix $\rho_{ch}=\frac{e^{-\beta H_{ch}}}{Z}$, where
$\beta=\frac{1}{k_BT}$, $k_B$ is the Boltzman constant and
$Z=tr(e^{-\beta H_{ch}})$ is the partition function. So the
initial state of system is
\begin{equation}\label{15}
\rho(0)=\ket{\psi^-}\bra{\psi^-}\otimes\frac{e^{-\beta
H_{ch}}}{Z}.
\end{equation}
Employing the system parameters as before, we repeat the
calculations for this new initial state and the results have been
shown in Figs 7 and 8 for $\Delta=1$ and $\Delta=-1$,
respectively. Two considerations are in order at this stage.
Firstly, at hight temperatures, thermal fluctuations suppress the
quantum correlations and hence the entanglement vanishes and also,
$t_{opt}$ becomes so large. Secondly, the presence of DMI amplify
the quantum correlations, so by increasing $D$ we can obtain
nonzero entanglement at larger temperatures. For instance Fig 8(a)
shows that for the case of $\Delta=-1$ the entanglement can be
exist at higher temperatures in the presence of DMI while it is
zero for all temperature in the absence of DMI.  Also, the speed
of transfer improves as $D$ increases.
\subsection{Information speed}
As mentioned before, the DMI imposes desirable effects on the
speed of transmission. In order to better clarify, we compare
$t_{opt}$ with $\emph{v}^{-1}$ (which $\emph{v}$ is the spin-wave
velocity) which is obtained with the aid of field theoretic
techniques \cite{field}. Note that, the later get the qualitative
behavior of correlations in the thermodynamic limit while the
former is calculated for very short chain ($N=8$). Also, in the
field theoretic techniques the dynamical correlations are computed
for the ground state of the system while in our problem it is not
the case. Despite difference, we still can use some well-known
results of the field theoretic techniques. For instance, in our
modified XXZ model(XXZ model equipped with DMI) and for the range
$-1<\Delta<1$, the spin-spin correlation function in the
asymptotic thermodynamical limit  have the following form
\cite{field}
\begin{eqnarray}\label{17}
\langle{\sigma^\mu_l(t)}{\sigma^\mu_{l+n}}&\rangle\sim&(-1)^n[n^2-\emph{v}^2t^2]^{-(1/2)\eta_\mu},\\
\eta_x=\eta_y&=&\eta_z^{-1}=1-\frac{\gamma}{\pi}\nonumber,
\end{eqnarray}
where
\begin{equation}\label{18}
\Delta=\cos{(\gamma)},
\end{equation}
and the propagation velocity of excitation in the chain can be
written as
\begin{equation}\label{19}
\emph{v}\sim\frac{\pi\tilde{J}\sin{(\gamma)}}{\gamma}.
\end{equation}
This velocity is proportional to the strength of DMI ($D$), which
is included in $\tilde J$, and also it depends on the size of
$\Delta$. According to the Ref. \cite{field}, $\emph v\,$ refers
to the velocity of propagation of the correlations which is
related to the entanglement transmission speed. Fig. 9 shows
$\emph{v}^{-1}$, which is obtained from Eq. (\ref{19}), and
$t_{opt}$, which is calculated in previous section, in terms of
$D$ for two different values of $\Delta$. As we can see both
quantities have similar behavior qualitatively, such that they
descent with increasing of $D$. As this figure illustrates,
despite of the field theory calculation for spin-wave propagation
time ($\emph v ^{-1}$), the curves of $t_{opt}$ cross each other,
indeed at the cross point the effect of the phase factor ($\phi$)
becomes considerable. As a consequence, for large values of $D$
the speed of information transmission raises up and ultimately
reaches the asymptotic value. Faster dynamics in large values of
$D$ stems to the entanglement enhancement of the channel in this
region.
\section{Conclusions}
In this paper we examined a XXZ Heisenberg chain equipped with the
Dzyaloshinski-Moriya interaction (DMI) as a quantum channel for
investigation of the entanglement transfer properties. We had
shown that the presence of DMI enhances the amount of coupling
constant, J, and also imposes a phase factor on the state of the
channel. In order to clarifying the role of DMI, we trace the
effects of the new coupling constant ($\tilde J$) and phase factor
($\phi$) for a wide range of the anisotropy parameter, $\Delta$,
separately. Indeed, increasing $\tilde J$ with $D$ leads to
increase the strength of spin-spin correlations and ultimately
improves the amount of the entanglement. For the case $\Delta<0$
the effects of $\phi$ on the entanglement properties of
transmission is more desirable. In contrast, for the case
$\Delta>0$ increasing $\phi$, individually, destroys the
entanglement of the channel. The effects of $\tilde J$ dominate
for large values of $D$ and hence the channel becomes more
efficient for all values of anisotropy parameter in the region
$-2\leq\Delta\leq2$. We calculated the entanglement properties for
the ground state and the thermal state of the chain, separately.
Our results show that the amount of entanglement and the speed of
transmission increase as $D$ increases. Also, we show that the
entanglement can be exist at higher temperatures as $D$ increases
and hence the transmission channel could be work more efficiently
at higher temperatures.
\appendix
\section{Analytical Calculation for N=4}
In this appendix, we give analytical calculation for the XXX and
XXZ chains with $N=4$, separately. For XXX chain ($\Delta=0$), the
initial state $\ket{\psi(0)}$ in the basis of the Hamiltonian
(\ref{5}) for $N=3$ can be written as \footnote{since the spin
$0^\prime$ does not coupled to the chain it remains untouched
during the evolution of the system. Thus, unitary transformation
affects only on the sites (0, 1, 2).}
\begin{eqnarray}
\ket{\psi(0)}&=&\frac{e^{-i\phi}}{2}\big{[}\ket{0}_{0^\prime}\otimes(\frac{e^{2i\phi}}{\sqrt{2}}\ket{\psi_2}+
\frac{1}{2}(\sqrt{2}+e^{2i\phi})\ket{\psi_5}+\frac{1}{2}(-\sqrt{2}+e^{2i\phi})\ket{\psi_7})\nonumber\\
&+&\ket{1}_{0^\prime}\otimes(-\frac{1}{\sqrt{2}}\ket{\psi_3}+\frac{1}{2}(1+\sqrt{2}e^{2i\phi})\ket{\psi_6}+
\frac{1}{2}(1-\sqrt{2}e^{2i\phi})\ket{\psi_8})\big{]},
\end{eqnarray}
where
\begin{eqnarray}
\ket{\psi_2}&=&\frac{1}{\sqrt{2}}(-\ket{011}+\ket{110}),\nonumber\\
\ket{\psi_3}&=&\frac{1}{\sqrt{2}}(-\ket{001}+\ket{100}),\nonumber\\
\ket{\psi_5}&=&\frac{1}{2}(\ket{011}-\sqrt{2}\ket{101}+\ket{110}),\nonumber\\
\ket{\psi_6}&=&\frac{1}{2}(\ket{001}-\sqrt{2}\ket{010}+\ket{100}),\\
\ket{\psi_7}&=&\frac{1}{2}(\ket{011}+\sqrt{2}\ket{101}+\ket{110}),\nonumber\\
\ket{\psi_8}&=&\frac{1}{2}(\ket{001}+\sqrt{2}\ket{010}+\ket{100}),\nonumber
\end{eqnarray}
are the relevant eigenstates of $H_3(\Delta=0)$ and the
corresponding eigenvalues are $E_2=E_3=0$, $E_5=E_6=-\xi$ and
$E_7=E_8=\xi$, here we define $\xi=2\sqrt{2}\tilde{J}$. The state
of system at later times can be obtained as
\begin{eqnarray}
\ket{\psi(t)}&=&\frac{e^{-i\phi}}{2}[\ket{0}_{0^\prime}\otimes(\frac{e^{2i\phi}}{\sqrt{2}}\ket{\psi_2}+
\frac{1}{2}(\sqrt{2}+e^{2i\phi})e^{i\xi t}\ket{\psi_5}\nonumber\\
&+& \frac{1}{2}(-\sqrt{2}+e^{2i\phi})e^{-i\xi t}\ket{\psi_7})+
\ket{1}_{0^\prime}\otimes(-\frac{1}{\sqrt{2}}\ket{\psi_3}\nonumber\\
&+&\frac{1}{2}(1+\sqrt{2}e^{2i\phi})e^{i\xi
t}\ket{\psi_6}+\frac{1}{2}(1-\sqrt{2}e^{2i\phi})e^{-i\xi
t}\ket{\psi_8})].
\end{eqnarray}
The corresponding reduced density matrix, $\rho_{0'2}(t)$ is in
the form of Eq. (\ref{10}) with the following components
\begin{eqnarray}
a&=&\frac{1}{16}\big{|}e^{2i\phi}(1+\cos{(\xi t)})+i\sqrt{2}\sin{(\xi t)}\big{|}^2,\nonumber\\
x&=&\frac{1}{16}\big{|}e^{2i\phi}(-1+\cos{(\xi t)})+i\sqrt{2}\sin{(\xi t)}\big{|}^2,\nonumber\\
&+&\frac{1}{8}\big{|}\sqrt{2}\cos{(\xi t)}+ie^{2i\phi}\sin{(\xi t)}\big{|}^2,\nonumber\\
y&=&\frac{1}{8}\big{|}\sqrt{2}e^{2i\phi}\cos{(\xi t)}+i\sin{(\xi t)}\big{|}^2,\\
&+&\frac{1}{16}\big{|}-1+\cos{(\xi t)}+i\sqrt{2}e^{2i\phi}\sin{(\xi t)}\big{|}^2,\nonumber\\
b&=&\frac{1}{16}\big{|}1+\cos{(\xi t)}+i\sqrt{2}e^{2i\phi}\sin{(\xi t)}\big{|}^2,\nonumber\\
z&=&\frac{1}{4}(-1+\cos{(\xi t)}).\nonumber
\end{eqnarray}
Therefore the concurrence could be computed as
\begin{eqnarray}\label{23}
C(t)&=&\frac{1}{8}\big{(}4\big{|}-1+\cos{(\xi t)}\big{|}-\big{|}e^{2i\phi}(1+\cos{(\xi t)})\nonumber\\
&+&i\sqrt{2}\sin{(\xi t)}\big{|}\times\big{|}1+\cos{(\xi t)}\nonumber\\
&+&i\sqrt{2}e^{2i\phi}\sin{(\xi t)}\big{|}\big{)}.
\end{eqnarray}
Following the same procedure for XXZ chain ($\Delta\neq0$), the
initial state $\ket{\psi(0)}$ in the basis of corresponding
Hamiltonian is
\begin{eqnarray}
\ket{\psi(0)}&=&\frac{e^{-i\phi}}{2}\big{[}\ket{0}_{0^\prime}\otimes(\frac{e^{2i\phi}}{\sqrt{2}}\ket{\psi_2}+
\frac{\alpha+e^{2i\phi}}{\sqrt{2+\alpha^2}}\ket{\psi_5}+\frac{-\beta+e^{2i\phi}}{\sqrt{2+\beta^2}}\ket{\psi_7})\nonumber\\
&+&\ket{1}_{0^\prime}\otimes(-\frac{1}{\sqrt{2}}\ket{\psi_3}+\frac{1+\alpha
e^{2i\phi}}{\sqrt{2+\alpha^2}}\ket{\psi_6}+\frac{1-\beta
e^{2i\phi}}{\sqrt{2+\beta^2}}\ket{\psi_8})\big{]},
\end{eqnarray}
where
$\alpha=\frac{\Delta+\sqrt{8\tilde{J}^2+\Delta^2}}{2\tilde{J}}$,
$\beta=-\frac{\Delta-\sqrt{8\tilde{J}^2+\Delta^2}}{2\tilde{J}}$
and
\begin{eqnarray}
\ket{\psi_2}&=&\frac{1}{\sqrt{2}}(-\ket{011}+\ket{110}),\nonumber\\
\ket{\psi_3}&=&\frac{1}{\sqrt{2}}(-\ket{001}+\ket{100}),\nonumber\\
\ket{\psi_5}&=&\frac{1}{\sqrt{2+\alpha^2}}(\ket{011}-\alpha\ket{101}+\ket{110}),\nonumber\\
\ket{\psi_6}&=&\frac{1}{\sqrt{2+\alpha^2}}(\ket{001}-\alpha\ket{010}+\ket{100}),\\
\ket{\psi_7}&=&\frac{1}{\sqrt{2+\beta^2}}(\ket{011}+\beta\ket{101}+\ket{110}),\nonumber\\
\ket{\psi_8}&=&\frac{1}{\sqrt{2+\beta^2}}(\ket{001}+\beta\ket{010}+\ket{100}),\nonumber
\end{eqnarray}
are the relevant eigenstates of $H_3(\Delta\neq0)$ and the
corresponding eigenvalues are $E_2=E_3=0$,
$E_5=E_6=-2\tilde{J}\alpha$ and $E_7=E_8=2\tilde{J}\beta$.
Therefore, the initial state, $\ket{\psi(0)}$ evolves to the state
\begin{eqnarray}
\ket{\psi(t)}&=&\frac{e^{-i\phi}}{2}\big{[}\ket{0}_{0^\prime}\otimes(\frac{e^{2i\phi}}{\sqrt{2}}\ket{\psi_2}+
\frac{(\alpha+e^{2i\phi})e^{-iE_5t}}{\sqrt{2+\alpha^2}}\ket{\psi_5}\nonumber\\
&+&\frac{(-\beta+e^{2i\phi})e^{-iE_7t}}{\sqrt{2+\beta^2}}\ket{\psi_7})
+\ket{1}_{0^\prime}\otimes(-\frac{1}{\sqrt{2}}\ket{\psi_3}\nonumber\\
&+&\frac{(1+\alpha
e^{2i\phi})e^{-iE_6t}}{\sqrt{2+\alpha^2}}\ket{\psi_6}+\frac{(1-\beta
e^{2i\phi})e^{-iE_8t}}{\sqrt{2+\beta^2}}\ket{\psi_8})\big{]}.
\end{eqnarray}
The corresponding reduced density matrix, $\rho_{0'2}(t)$ is in
the form of Eq. (\ref{10}) with the following components
\begin{eqnarray}
a&=&\frac{1}{4}|\frac{e^{2 i \tilde{J} t \alpha } (\alpha +e^{2 i
\phi})}{\alpha ^2+2}+\frac{1}{2} e^{2 i \phi }+\frac{e^{-2 i
\tilde{J} t\beta } (e^{2 i \phi }-\beta )}{\beta^2+2}|^2,\nonumber\\
x&=&\frac{1}{4}\big{|}-\frac{1}{2}e^{2i\phi}+\frac{(\alpha+e^{2i\phi})e^{2i\alpha \tilde{J} t}}{2+\alpha^2}+
\frac{(-\beta+e^{2i\phi})e^{-2i\beta \tilde{J} t}}{2+\beta^2}\big{|}^2\nonumber\\
&+&\frac{1}{4}\big{|}-\frac{\alpha(\alpha+e^{2i\phi})e^{2i\alpha \tilde{J} t}}{2+\alpha^2}+
\frac{\beta(-\beta+e^{2i\phi})e^{-2i\beta \tilde{J} t}}{2+\beta^2}\big{|}^2,\nonumber\\
y&=&\frac{1}{4}\big{(}\big{|}-\frac{1}{2}+\frac{(1+\alpha
e^{2i\phi})e^{2i\alpha \tilde{J} t}}{2+\alpha^2}+\frac{(1-\beta
e^{2i\phi})e^{-2i\beta \tilde{J} t}}{2+\beta^2}\big{|}^2\nonumber\\
&+&\big{|}-\frac{\alpha(1+\alpha e^{2i\phi})e^{2i\alpha \tilde{J}
t}}{2+\alpha^2}+\frac{\beta(1-\beta e^{2i\phi})e^{-2i\beta \tilde{J} t}}{2+\beta^2}\big{|}^2\big{)},\\
b&=&\frac{1}{4}\big{|}\frac{1}{2}+\frac{(1+\alpha e^{2i\phi})e^{2i\alpha \tilde{J} t}}{2+\alpha^2}+
\frac{(1-\beta e^{2i\phi})e^{-2i\beta \tilde{J} t}}{2+\beta^2}\big{|}^2,\nonumber\\
z&=&\frac{1}{4} ((\frac{\beta  (-\beta +e^{-2 i \phi })
e^{2i\tilde{J} t \beta }}{\beta ^2+2}-\frac{\alpha (\alpha +e^{-2
i \phi }) e^{-2 i \tilde{J} t \alpha}}{\alpha ^2+2})\nonumber\\
    &\times&(\frac{(1+\alpha  e^{2 i \phi }) e^{2 i \tilde{J} t \alpha }}{\alpha ^2+2}+
    \frac{(1-\beta  e^{2 i \phi }) e^{-2 i \tilde{J} t \beta }}{\beta ^2+2}-\frac{1}{2})\nonumber\\
   &+&(\frac{(\alpha +e^{-2 i \phi }) e^{-2 i \tilde{J} t \alpha }}{\alpha ^2+2}+
   \frac{(-\beta +e^{-2 i \phi }) e^{2 i \tilde{J} t \beta }}{\beta ^2+2}-\frac{1}{2} e^{-2 i \phi })\nonumber\\
   &\times& (\frac{\beta  (1-\beta  e^{2 i \phi }) e^{-2 i J t \beta }}{\beta ^2+2}-\frac{\alpha  (1+\alpha
   e^{2 i \phi }) e^{2 i \tilde{J} t \alpha }}{\alpha ^2+2})),\nonumber
\end{eqnarray}
and hence
\begin{eqnarray}\label{25}
C(t)&=&\half (|(\frac{e^{2 i \tilde{J} t \beta } (e^{-2 i \phi
}-\beta ) \beta }{\beta ^2+2}-\frac{e^{-2 i \tilde{J} t \alpha }
\alpha (\alpha +e^{-2 i \phi
   })}{\alpha ^2+2})\nonumber\\
   &\times&(\frac{e^{2 i \tilde{J} t \alpha } (e^{2 i \phi } \alpha +1)}{\alpha ^2+2}+
   \frac{e^{-2 i \tilde{J} t \beta } (1-e^{2 i \phi } \beta
   )}{\beta ^2+2}-\frac{1}{2})\nonumber\\
   &+&(\frac{e^{-2 i \tilde{J} t \alpha } (\alpha +e^{-2 i \phi })}{\alpha ^2+2}-\frac{1}{2} e^{-2 i \phi }+
   \frac{e^{2 i \tilde{J} t \beta
   } (e^{-2 i \phi }-\beta )}{\beta ^2+2})\\
    &\times&(\frac{e^{-2 i \tilde{J} t \beta } \beta  (1-e^{2 i \phi } \beta )}{\beta ^2+2}-
    \frac{e^{2 i \tilde{J} t \alpha }
   \alpha  (e^{2 i \phi } \alpha +1)}{\alpha ^2+2})|\nonumber\\
   &-&|\frac{e^{2 i \tilde{J} t \alpha } (\alpha +e^{2 i \phi })}{\alpha
   ^2+2}+\frac{1}{2} e^{2 i \phi }+\frac{e^{-2 i \tilde{J} t \beta } (e^{2 i \phi }-\beta )}{\beta ^2+2}|\nonumber\\
   &\times&|\frac{e^{2 i \tilde{J} t \alpha } (e^{2 i \phi } \alpha
   +1)}{\alpha ^2+2}+\frac{e^{-2 i \tilde{J} t \beta } (1-e^{2 i \phi } \beta )}{\beta
   ^2+2}+\frac{1}{2}|).\nonumber
\end{eqnarray}
%%%%%%%%%%%%%%%%%%%%%%%%%%%%%%%%%%%%%%%%%%%%%%%%%%%%%%%%%%%%%%%%%%%%%%%%%%%%%%%%%%%%%%%%%%%%%%

%Figures%%%%%%%%%%%%%%%%%%%%%%%%%%%%%%%%%%%%%%%%%%%%%%%%%%%%%%%%%%%%%%%
\newpage
\begin{figure}
\epsfxsize=12cm\ \centerline{\hspace{0cm}\epsffile{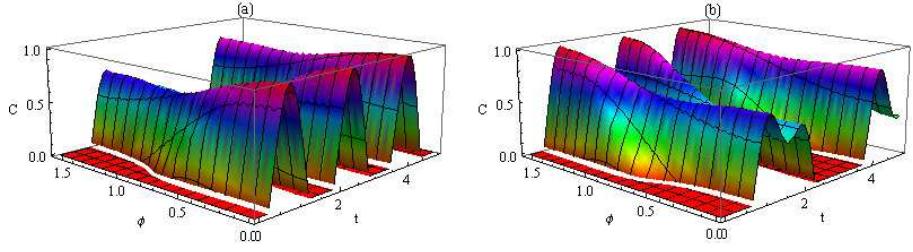}} \
\caption{(Color online) $C(\rho_{0'4})$ vs. $\phi$ and $t$, for
(a) $\Delta=0.9$ and (b) $\Delta=-0.9$. Here the length of the
chain is chosen to be $N=4$.}
\end{figure}
\begin{figure}
\epsfxsize=12cm\ \centerline{\hspace{0cm}\epsffile{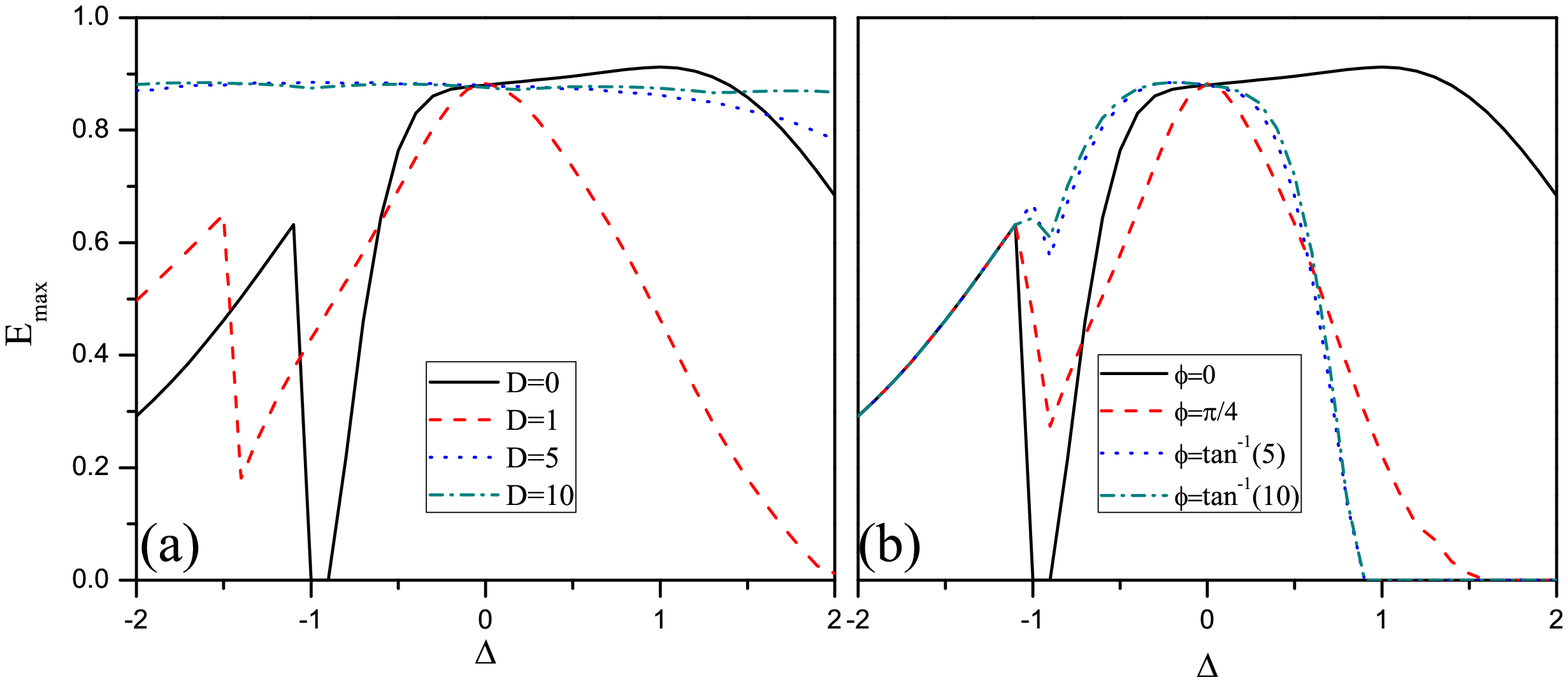}} \
\caption{(Color online) $E_{max}$ as a function of $\Delta$, for
(a) different values of $D$ and (b) corresponding $\phi$.}
\end{figure}
\begin{figure}
\epsfxsize=12cm\ \centerline{\hspace{0cm}\epsffile{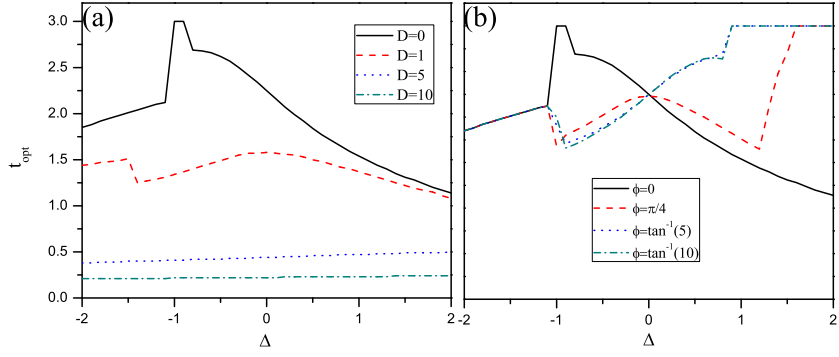}} \
\caption{(Color online) $t_{opt}$ vs. $\Delta$, for (a) different
values of $D$ and (b) corresponding $\phi$.}
\end{figure}
\begin{figure}
\epsfxsize=10cm\ \centerline{\hspace{0cm}\epsffile{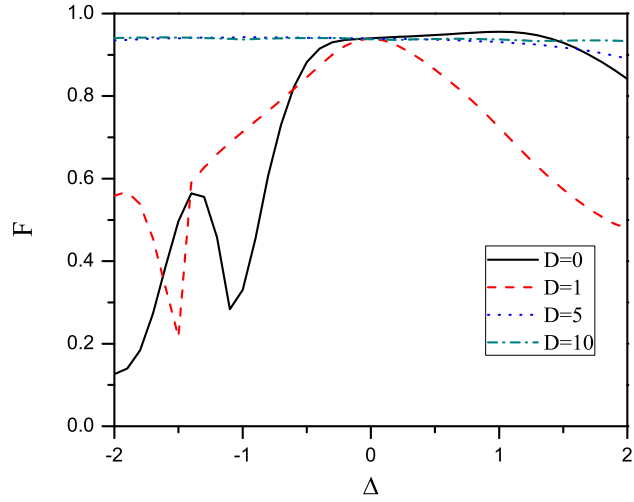}} \
\caption{Singlet fraction $F$ vs. $\Delta$ at $t_{opt}$ for
different values of $D$.}
\end{figure}
\begin{figure}
\epsfxsize=12cm\ \centerline{\hspace{0cm}\epsffile{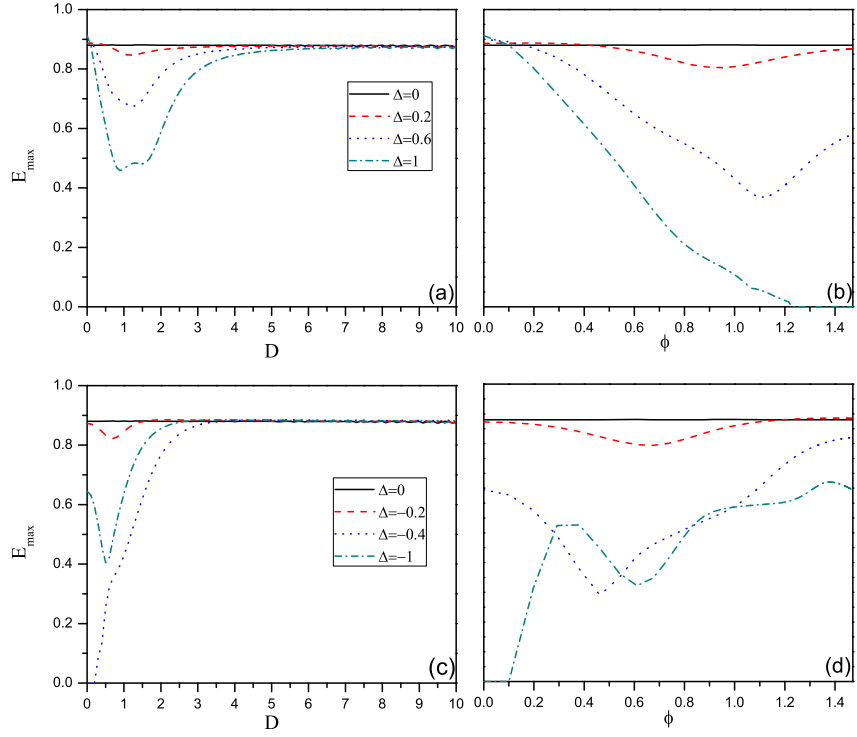}} \
\caption{(Color online) First maximum entanglement vs. $D$ and
corresponding $\phi$. Graphs (a) and (b) refer to the case
$\Delta\geq 0$ and (c) and (d) refer to the case $\Delta\leq 0$.}
\end{figure}
\begin{figure}
\epsfxsize=12cm\ \centerline{\hspace{0cm}\epsffile{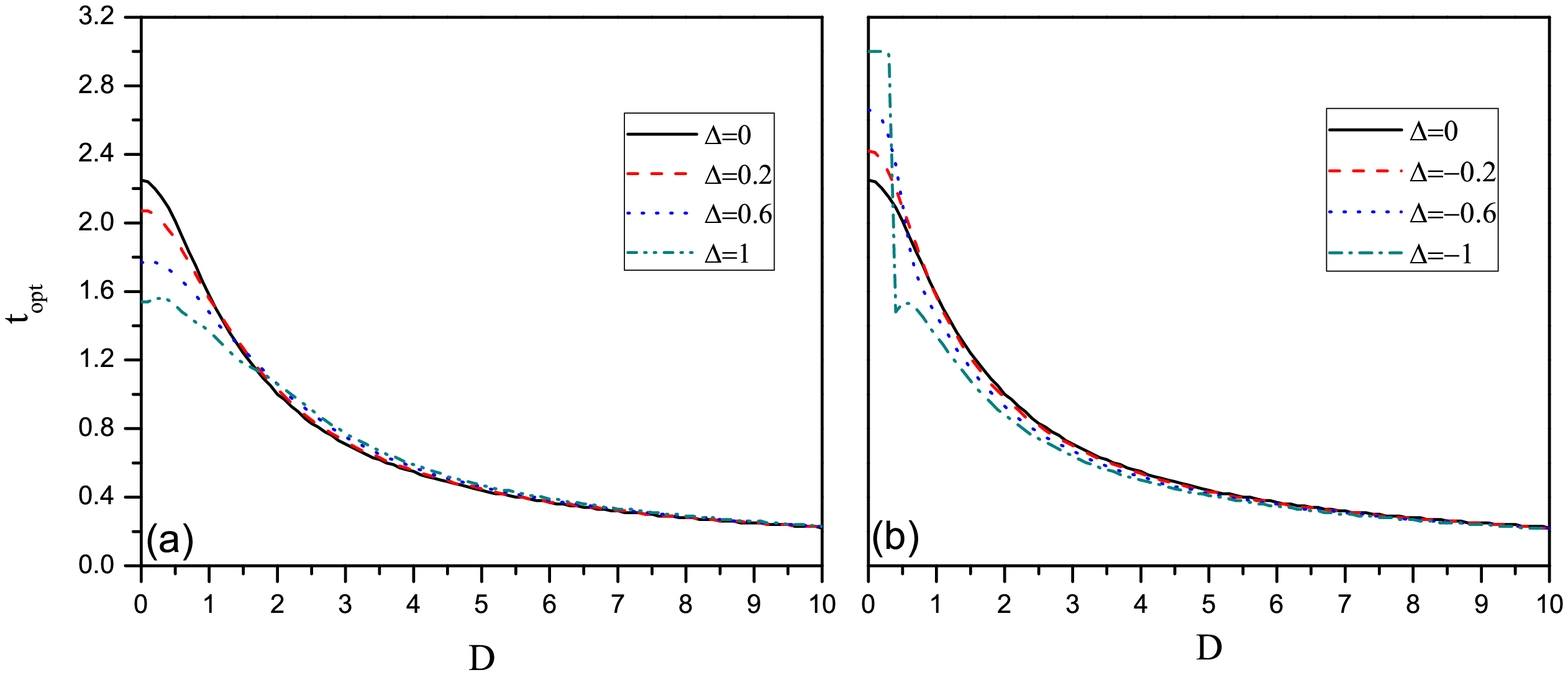}} \
\caption{(Color online) $t_{opt}$ in terms of $D$, for different
values of (a) $\Delta\geq 0$ and (b) $\Delta\leq0$.}
\end{figure}
\begin{figure}
\epsfxsize=12cm\ \centerline{\hspace{0cm}\epsffile{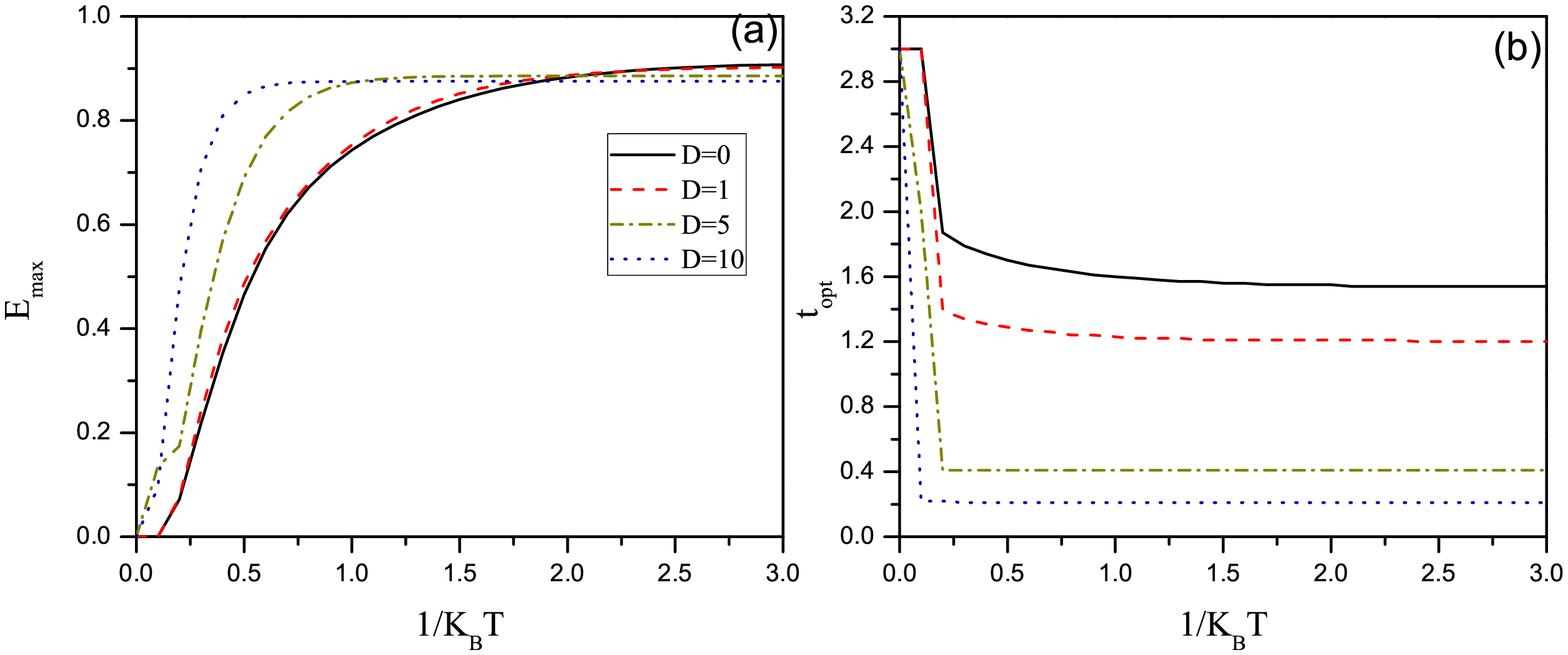}} \
\caption{(Color online) (a) $E_{max}$ and (b) $t_{opt}$ in terms
of inverse temperature for  different values of $D$ at
$\Delta=1$.}
\end{figure}
\begin{figure}
\epsfxsize=12cm\ \centerline{\hspace{0cm}\epsffile{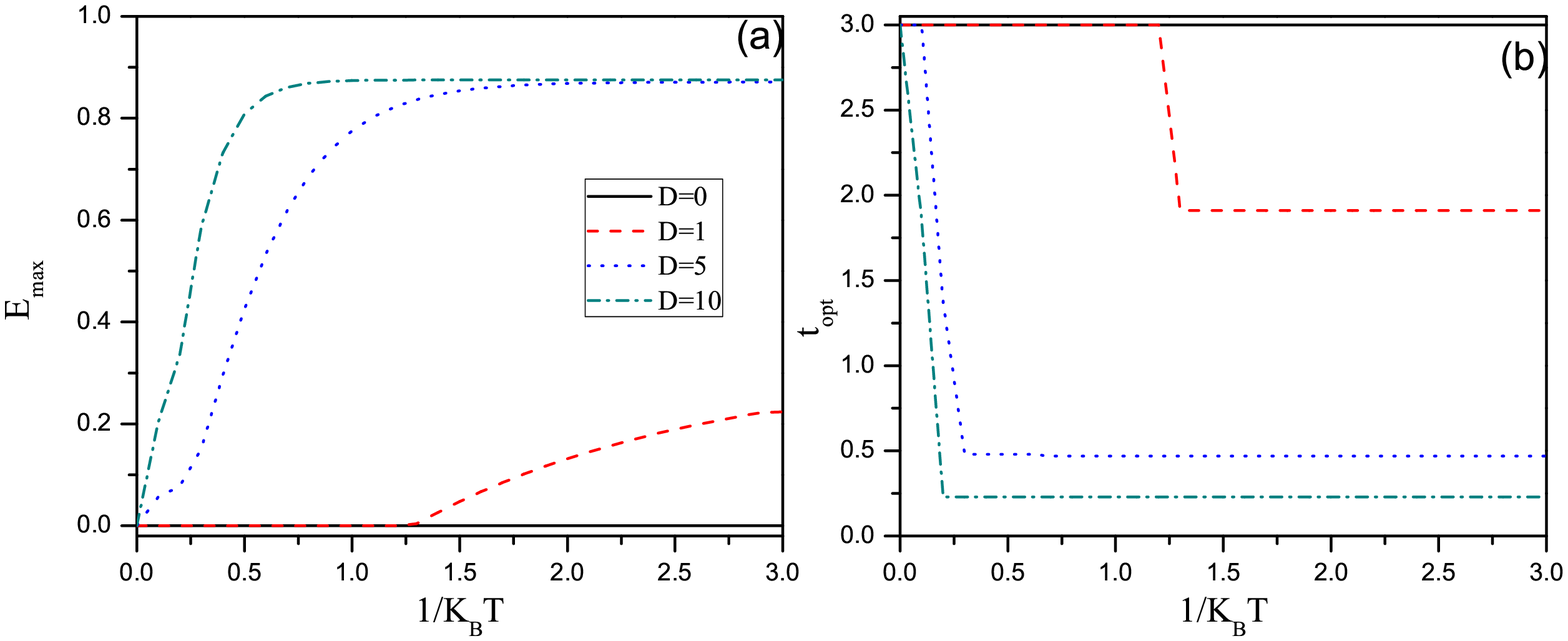}} \
\caption{(Color online) $E_{max}$ and (b) $t_{opt}$ in terms of
inverse temperature for  different values of $D$ at $\Delta=-1$.}
\end{figure}
\begin{figure}
\epsfxsize=10cm\ \centerline{\hspace{0cm}\epsffile{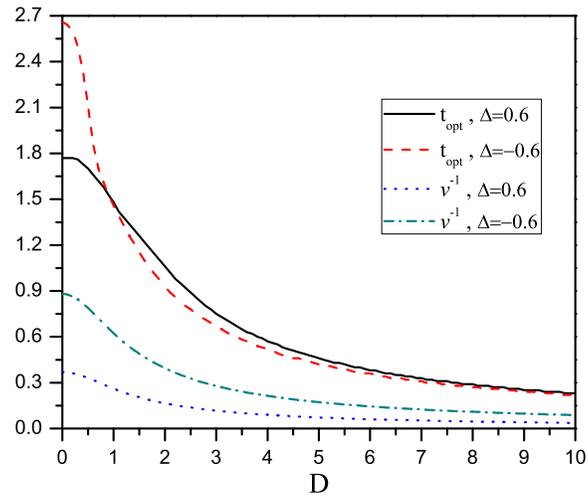}} \
\caption{(Color online) $\emph{v}^{-1}$ (for an infinite chain)
and $t_{opt}$ (for the chain of length N=8) vs. D for the case of
$\Delta=0.6$ and $\Delta=-0.6$.}
\end{figure}
\end{document}